\documentstyle[aps,epsfig]{revtex}                  
\input epsf

\begin{document}
\pagestyle{empty}                                      

\draft
\vfill
\title{
Regular charged black hole with a charged de Sitter core}
\author{W. F. Kao} %
\address{Institute of Physics, Chiao Tung University,
Hsinchu, 300, Taiwan}
\date{\today}
%
%
\vfill
\maketitle
\begin{abstract}
We show that there is a class of regular charged black hole with a charged de Sitter core similar to the de Sitter-Schwarzschild black hole.
One can show that the total energy momentum tensor for the static and spherically symmetric photon field can be made conserved even the Coulomb like energy momentum tensor for photon is not conserved alone.
Possible impact and the numerical solutions of the charged black hole will be shown in this paper. 
\end{abstract}
%
%
%
\pagestyle{plain}
\section{Introduction}

The Penrose cosmic censorship states that: if singularities predicted by General Relativity \cite{H-E,Senov98} occur, they must be dressed by event horizons \cite{Wald97,JMagli99}.
Hence, the search of the regular black hole solutions has been a focus of research interests lately \cite{reg1,reg2,reg3,reg4}. 
In fact, the study of global regularity of black hole solutions is important for the understanding of the final state of the gravitational collapse from some initially regular configurations.

Earlier work on regular black hole models can be found in References \cite{Bardeen68,Ayon93,Borde94,BarrFrolov96,MMPSenovilla96,CaboAyon97}.
These regular solutions are referred to as ``Bardeen black holes'' \cite{Bardeen68,Borde97}. 
In addition, regular black hole solutions to Einstein equations with various physical sources were reported in \cite{reg1,AyonGarcia98,Magli97}. 
The other approaches to avoid the singularities are the search of some more generalized theories, for example, the string/M--theory  \cite{Tseytlin95}.

The Einstein equation reads
\begin{equation}
G_{\mu\nu} =- 8 \pi T_{\mu\nu}
\end{equation}
where $G_{\mu\nu}\equiv {1 \over 2} g_{\mu \nu}R - R_{\mu \nu} $ is the Einstein tensor and
$T_{\mu\nu}$ is the matter stress-energy tensor.
For example, $T_{\mu \nu}= (\Lambda/8 \pi) g_{\mu \nu}$ is the contribution from a cosmological constant $\Lambda$.
In the absence of the $T_{\mu\nu}$ associated with the other matter, $\Lambda$ must be a constant.
This follows from the Bianchi identity: $D_\mu G^{\mu\nu}=0$.

There have been activities trying to treat the cosmological constant $\Lambda$ as a dynamical quantity \cite{bahcall,s1,s2,s3,s5,s7,s9,s10}. 
The vacuum stress-energy tensor has the form $T_{\mu\nu}=<\rho_{vac}>g_{\mu\nu}$ in the semi-classical approach of the quantum field theory in curved space. In this approach, the cosmological term becomes 
$\Lambda= 8 \pi <\rho_{vac}>$ which has been studied for more than two decades \cite{adler,weinberg,overduin,gliner,zeldovich,andrei,us75,dolgov,petrov}.

Note that the de Sitter-Schwarzschild solution found in Ref. \cite{reg5} is a black hole solution with a de Sitter core near $r=0$.
We will generalize this solution to the charged regular solution by assuming that the matter coupled to the gravitational system carries charge.
We assume that the charged material has a uniform charge/mass ratio all over.
Therefore, the $r$-dependence of the charged density $d_e(r)$ is the same as the
$r$-dependence of the mass density $\rho(r)$ for the charged material.
We will show that a regular charged black hole solution exists with a charged de Sitter core.
We will also study the properties of this regular charged solution and its possible impact.

\section{Spherically collapsing star}

In the static and spherically symmetric space, the line element can be written in the form \cite{tolman}
\begin{equation}
ds^2 = -\exp [2A(r)] dt^2 + \exp[2B(r)] dr^2 + r^2 d\Omega
\end{equation}
where $d\Omega$ is the solid angle.
One can write $\exp [-2B] = 1- 2N(r)/r$ and show that the Einstein equation 
$G^t_t = -2T^t_t= - 2 \rho(r)$ reads
\begin{equation}
N'=4 \pi r^2 \rho . \label{N-eq}
\end{equation}
It is known that the energy momentum tensor of the static and spherically symmetric star takes the following form 
\begin{eqnarray}
T^\theta_{\, \theta}&=&T^\varphi_{\, \varphi}=p(r).
\end{eqnarray}
with $p$ signifies the pressure contribution.
For simplicity we will work on the case where
\begin{eqnarray}
T^t_{\, t}&=& T^r_{\, r}= \rho(r) , 
\end{eqnarray}
which is classified as the type [(II),(II)] field according to the classification of Serge \cite{petrov}.
The $tt$ and $rr$ components of the Einstein equation give immediately that
$A+B=0$ once we choose the proportional constant to be zero.
In addition, the conservation of energy momentum $D_\mu T^{\mu \nu}=0$ requires that 
\begin{equation}
p= (r/2) \rho' +  \rho . 
\end{equation}
This equation tells us what the pressure term $p$ should look like with a given energy density $\rho$ in order to stabilize the space induced by the Serge's type [(II),(II)] field we are interested.
For example, $p= \Lambda /2 - Q^2/r^4, \:$ if $ \: 
\rho= Q^2/r^4 +\Lambda/2 $.
In addition, one has  $N=M- Q^2/r +\Lambda r^3/6$ for constant $M$, $Q$ and $\Lambda$ according to equation (\ref{N-eq}).
Note that this is the Reissner-Nordstrom solution with a cosmological constant term.

Note that the $U(1)$ gauge field will provide a Lagrangian of the form
\begin{equation}
{\cal L}_F= - F^2/8 \pi.
\end{equation}
This will contribute a traceless energy momentum tensor of the form
\begin{equation}
T_e^{\mu \nu} = g^{\mu \nu} F^2 / 16 \pi- F^\mu_\alpha F^{\nu \alpha}/4 \pi
\end{equation}
to the Einstein equation.
In addition, the $U_1$ electric charge conservation $D_\mu F^{\mu \nu}=4 \pi J^\nu$ with a static source $J^\mu = (d_e(r), {\bf 0})$ gives the system a static Coulomb potential $A_\mu =( \phi(r), {\bf 0})$ in the Coulomb's gauge.
One can show that the electric field is $E_r =F_{tr}= Q(r)/r^2$ for this spherically symmetric source.
Here $Q(r)= 4 \pi \int_0^r r'^2 dr' d_e(r')$ is the total charge surrounded by the $r$-sphere.
Hence the energy momentum tensor for this charged material can be shown to be 
\begin{eqnarray}
\rho_e &\equiv& T_{e\,t} ^{\: t} = T_{e\,r} ^{\: r} = Q^2/8 \pi r^4 \\
p_e &\equiv& T_{e\,\theta} ^{\: \theta}= T_{e\,\varphi} ^{\: \varphi}= - \rho_e .
\label{cmpre}
\end{eqnarray}
Here we purposely denote the charge density as $d_e$ because the mass density happens to be the energy density of the matter sector while the energy density of the photon sector $\rho_e$ is not the same as the charge density $d_e$.

Note also that one should have 
\begin{equation}
 p_e = r \partial_r \rho_e /2  + \rho_e 
\end{equation}
if the photon energy momentum tensor is conserved, $D_\mu T_e^{\mu \nu}=0$, by itself. 
But, this is true only if the total charge $Q$ is a constant everywhere.
Hence, the conservation of energy momentum tensor $T_e^{\mu \nu}$ is known to be in-consistent with the $Q(r)$-dependent static electric field given by
$E_r =Q(r)/r^2$ for any $r$-dependent $Q=Q(r)$.
In fact, the conservation law requires an $1/r^4$ energy density because
\begin{equation}
r \partial_r \rho_e + 2  \rho_e = -2 \rho_e$ for $\rho_e=-p_e = {\rm constant} \times 1/r^4
\end{equation}
for a the Coulomb's electric field with a constant $Q$.
This easy to verify since the operator $r \partial_r $ is nothing more than a power counting operator.
Hence the conservation equation 
\begin{equation}
r \partial_r \rho_e  = 2 p_e- 2 \rho_e = -4 \rho_e
\end{equation}
simply implies that $\rho_e \sim 1/r^4$.
This is a problem one has to deal with in working with the charged solutions for a regular distribution of charged material.
In fact, the resolution is to assign an additional pressure contribution to the matter sector.

Indeed, one can show that the total energy momentum conservation remain valid if the total pressure $p$ is given by
\begin{equation}
p = p_e +p_m = p_e + [\tilde{p}_m + QQ'/8 \pi r^3 ] 
\end{equation}
for the $r$-dependent $Q(r)$ given earlier.
Here $\tilde{p}_m = r \rho_m'/2 + \rho_m$ is the matter pressure in the absence of the $r$-dependent $Q$.
In other word, the system remains consistent if the matter pressure $p_m$ offers an extra contribution in order to counteract the contribution due to the $Q(r)$ term.
Note that we only show that a stable static system requires an additional pressure for stability.
Stable charged stars certainly can exit and this extra pressure simply signifies the contribution from a complicated interaction among the charged material.

Let's imagine a charged material distribution of the following form
\begin{eqnarray} \label{cmr}
\rho_m &=& 3 M_0 \exp [-r^3] / 4 \pi  , \\
\label{cme}
d_e &=& 3Q_0 \exp [-r^3] / 4 \pi.
\end{eqnarray}
One can show that the total mass and total charge measured at a distance $r$ read
$M(r)= M_0 ( 1- \exp [-r^3] \,)$ and
$Q(r)= Q_0 ( 1- \exp [-r^3] \,)$ for this charged material.
In fact, one can show that the energy of the whole system is
\begin{eqnarray}
\rho = \rho_m + \rho_e = 3M_0 \exp [-r^3 ] / 4 \pi  + [Q_0^2/8 \pi r^4] (1- \exp [ -r^3] \,)^2 . \label{pm}
\end{eqnarray}
Note that $M_0$ and $Q_0 \equiv q M_0$ denote the total mass and total charge measured at spatial infinity.
One has assumed that each charged matter $X$ carries mass $m_X$ and a fixed unit of charge $Ze$ such that $q = NZe/M_0 =Ze/m_X$.
Here $N \equiv M_0/m_X$ denotes the total number of the molecular and/or charged material $X$.

One can show that the matter pressure of this system is given by
\begin{equation}
p_m = 3M_0 \exp [-r^3 ] / 4 \pi - 9M_0 r^3 \exp [-r^3 ] / 8 \pi
+ 3Q_0^2 \exp [-r^3 ] (1- \exp [-r^3] ) / 8 \pi r
\label{cmprm}
\end{equation}
with the last term signifies the counter term from the charged sector.

One can imagine that charged material carrying mass quanta and charge quanta tries to form a regular black hole with a charged de Sitter (cdS) core.
If the uncharged matter can form a regular de Sitter-Schwarzschild black hole, the charged material should also be able to form a charged regular black hole.
Hence we will try to study the possibility of finding such kind of regular black hole solution.
Note that we have changed the scale $r$ to be dimensionless such that $r$ should be $r/r_0$ for a physical scale $r_0$ typically of the size of radius of the event horizon.
In this approach, the charged material is bounded mostly inside the $r_0$-sphere.

Note that the weak energy condition, $T_{\mu \nu}v^\mu v^\nu \ge 0$, for any time-like vector $v^\mu$ is violated as $r \to 0$ for this solution.
Indeed, one can show that
\begin{equation}
4 \pi (\rho -p) = Q^2/r^4+9M_0r^3\exp [-r^3]/2-3Q_0^2\exp [-r^3] (1- \exp [-r^3])/2r \ge 0
\end{equation}
Therefore, the weak energy condition also requires
$9M_0r^3 -Q_0^2r^2 \ge 0$ for all $r \ll 1$.
This inequality will break down for $r < Q_0^2/9M_0$. 
Hence the existence of such solution will require the existence of exotic matter \cite{time}.

One way out is to assume the charge distribution grows as $\exp [-r^l]$, namely,
$Q( r )= Q_0(1 - \exp [-r^l] )$ for some $l\ge 2$.
One can show that the inequality $l \ge 2$ is needed for a regular solution which demand all scale-invariant curvature-square terms like $R^2$ be finite everywhere.
Hence one can show that the weak energy condition reads
\begin{equation}
4 \pi (\rho -p) = Q^2/r^4+9M_0r^3\exp [-r^3]/2-lQ_0^2\exp [-r^l] (1- \exp [-r^l])/2r \ge 0
\end{equation}
Therefore, the weak energy condition becomes
$Q_0^2 r^{2l-4} +9M_0r^3/2 -lQ_0^2r^{l-1}/2 \ge 0$ when $r \ll 1$.
This inequality can be made valid for all $r \ll 1$ if $l=4$ and $M_0 \ge 4Q_0^2/9$.
Note that this is not the necessary condition for the weak energy condition.
Therefore, smaller charge accumulated around the matter distribution can save the weak energy condition from breaking-down.

\section{Black hole solution}
Note that the charged black hole of this type can be shown to give the following metric solution
\begin{equation}
1/g_{rr}=g^{rr}= 1-  2M_0 (1 - \exp [-r^3] )/r 
+ Q_0^2 (1 - \exp [-r^l] )^2 /r^2 - [2l Q_0^2/r] \int_0^r r'^{l-2} dr' \exp [-r'^l] (1 - \exp [-r^l] ). \label{grr}
\end{equation}
We will focus on the case where $l=4$ for the moment.
Note that the last term can be integrated to give
\begin{equation}
-[8Q_0^2/r] \{ \Gamma[7/4]/3 -  \Gamma [3/4, r^4] /4- ( \Gamma[7/4] -  \Gamma [3/4, 2r^4])/2^{3/4} \}
\end{equation}
described by the di-gamma function when $l=4$.
Note that the last term can be shown to give $-8Q_0^2r^6/7$ if $r \ll 1$.
It is $-2lQ_0^2r^{2l-2}/(2l-1)$ if $r \ll 1$ for arbitrary $l$.
In fact, if the horizon is located in region $r \ll 1$, the metric $g^{rr}$ reads
$1-2M_0r^3 - Q_0^2r^6/7$.
This equation is different from the Reissner-Nordstrom solution by the sign of the last term related to $Q_0^2$.
Hence there always exist two distinct event horizons for this regular charged black hole if the horizon exists deep inside the $r_0$-core, i.e. $r/r_0 \ll 1$.

But this is not the correct picture for a black hole.
Matter should reside mostly inside the black event horizon where $g_{rr}$ diverges.
Hence one expects that the radius of the event horizons should be greater than the typical size of the $r_0$-core.
Therefore, the above naive picture does not hold for the region where $r>1$.
In fact, numerical analysis indicates that there is an event horizon and a Cauchy horizon if $M_0 > 0.625$ in our normalized unit $Q_0=1/\sqrt{2}$.
Numerical results are plotted in Fig. I.


\begin{figure}
  \unitlength 1mm
   \begin{center}
      \begin{picture}(25,100)
\put(-60,20) {\epsfig{file=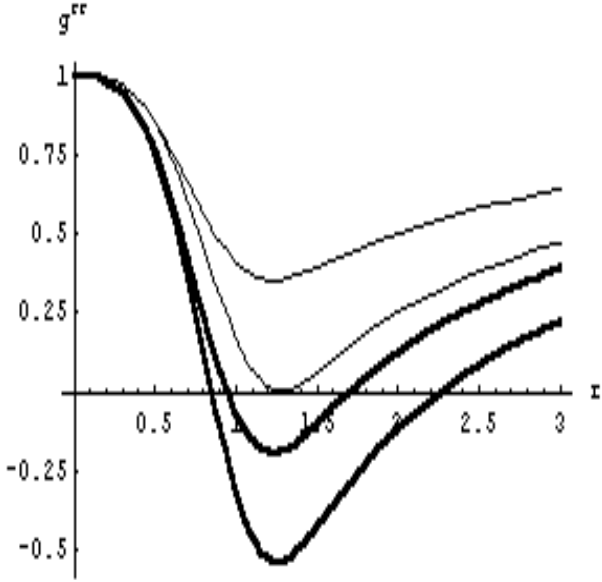,width=8cm,height=8cm}}
      \end{picture}
   \end{center}
\caption{The lower (upper) thick curve denotes the $g^{rr}$-$r$ curve for the case $M_0=1$ with (without) the last digamma term in Eq. (\ref{grr}).
The lower (upper) thin curve goes for the case $M_0=0.625$ with (without) the last digamma term.
The lower thin curve indicates the case where extreme charged black hole exists. 
}
\label{fig1}
\end{figure}

Figure I is plotted only for reference with the total charge set equal to $1/\sqrt{2}$ and mass $M_0$ set as $M_0=1, 0.625$.
The lower thick curve denotes the $g^{rr}$-$r$ curve for the case $M_0=1$ with  the last digamma term in Eq. (\ref{grr}).
The lower  thin curve goes for the case $M_0=0.625$ with the last digamma term in Eq. (\ref{grr}).
Note that the lower thin curve indicates the case where extreme charged black hole exists. 
These are plotted with $g^{rr}$ given by Eq. (\ref{grr}).
In addition, the upper thin/thick curves are the curves plotted only with $g^{rr}$ given by the first three terms in Eq. (\ref{grr}).
This result indicates that the contribution of the di-Gamma functions can not be ignored.
Note also that all curves approach $g^{rr} \to 1$ at spatial infinity.

It is known that the event horizon disappears for the Reissner-Nordstrom solution when $Q_0>M_0$.
The extreme black hole is, however, not present at the condition $Q_0=M_0$ in this case. 
In fact, the mass/charge distribution introduced here deforms the extreme condition into a very complicated condition.
We are, unfortunately, unable to solve algebraically for the exact expression of the $Q_0$-$M_0$ relation for the condition of the extreme black hole. 
The numerical study shows, however, that the physical behavior of the black hole horizons is similar to the Reissner-Nordstrom black hole.
Event horizon will disappear under a somewhat complicated $Q_0$-$M_0$ relation.
Note also that one can show that all curvature tensor $R^{\mu \nu}_{\:\alpha \beta} < \infty$ for the regular charged black hole solution we just obtained.
Therefore, this charged solution is indeed regular at $r=0$ since all $R^2$-invarinats are finite at $r=0$.

\section{conclusion}

In summary, we have shown that there is a class of regular charged black hole with a charged de Sitter core similar to the de Sitter-Schwarzschild black hole for the type [(2,2)] charged matter.
The associated conservation law of energy momentum tensor is quite different from the uncharged counterpart.
One can show that the energy momentum tensor associated with the static and spherically symmetric photon field is not conserved by itself.
The resolution is to assign an extra photon contribution to the pressure term associated with the matter field $\tilde{p}_m$.
Possible impact and the numerical solutions of the charged black hole are also shown in this paper. 
The structure of this cdS solution could be important to the physics of the gravitationally collapsing material.
In addition, a more general form of the matter distribution can also be shown to establish the stability of these models \cite{kao2}.
This is the first regular exact solution with a charged de Sitter core in general relativity based on a physical distribution of charged material \cite{reg5}.
It should be important to the understanding of the star evolution if the de Sitter-Schwarzschild solution is realized in nature.

It is known that the case when $l=3$ will lead to the demand of exotic matter that violates the weak energy condition when $r$ is small.
This problem can be resolved by assuming a smaller charge distribution function, e.g. assuming $l=4$.
We show explicitly that the weak energy condition remains valid for the choice of $l=4$ provided that $M_0$ is much larger than $Q_0^2$.
For simplicity, we only discuss the case where $l=4$ in most of this paper.
The generalization to arbitrary $l$ is straightforward. 

No-hair theorem states that all static black holes are characterized uniquely by mass and charge, and have the Reissner-Nordstrom form. 
Similar to the regular solution obtained in Ref. \cite{reg5}, our solution also appears to violate the no-hair theorem.
In fact, the existence of this sort of regular solution requires a mass and charge distribution of the form given by Eq.s (\ref{cmr}, \ref{cme}) along with a pressure given by Eq.s (\ref{cmprm}) and (\ref{cmpre}).
This is quite different from the usual vacuum solution we have before.
For example, the behavior of a test particle in-falling to the black hole will require additional knowledge about the constituent of the static $star$.
Numerical result also indicates that the size of event horizon is of the same order of the size that most of the matter resides, i.e. $r_H \sim O(1)$ while matter (charge) density decreases exponentially according to $\exp[-r^3]\: ( \exp[-r^4])$ .
In addition, the regular solution we just obtained approaches the R-N solution asymptotically.
Therefore, the only measurable quantities in asymptotic region are $Q_0$ and $M_0$.
Hence one can fairly say that the no-hair theorem holds softly for this solution too. This is because all information beyond the event horizon is not reachable from exterior observer.
Even it is not clear how one can arrange mass and charge the way we want in the core region, it is still a heuristic analysis to study regular models similar to the one studied in this paper.

\vskip 2cm

{\bf \large Acknowledgments}
The author thanks J.J. Tseng for helps in running Mathematica.
This work was supported in part by NSC under the contract number NSC89-2112-M009-043.

\end{document}